\documentclass[11pt]{article}
\usepackage{amsfonts}
\usepackage{amssymb}
\usepackage{amsmath}
\usepackage{amsthm}
\usepackage{amsfonts}
\usepackage{multicol}
\usepackage{amscd}
\usepackage{textcomp}
\usepackage{multirow}
\usepackage[english, activeacute]{babel}

\def\be{\begin{equation}}
\def\ee{\end{equation}}
\def\bea{\begin{eqnarray}}
\def\eea{\end{eqnarray}}

\newcommand{\sect}[1]{\setcounter{equation}{0}\section{#1}}
\newcommand{\subsect}[1]{\subsection{#1}}
\newcommand{\subsubsect}[1]{\subsubsection{#1}}

\def\otra{b}

\def\1{\'{\i}}

\def\>#1{{\mathbf#1}}

 \def\otra{b}

\begin{document}

\thispagestyle{empty}


\ 
\vspace{0.5cm}

\begin{center}

{\Large{\sc{Integrable H\'enon-Heiles Hamiltonians: 
\\
a Poisson algebra approach}} }

\end{center}

\medskip

\begin{center} \'Angel Ballesteros and Alfonso Blasco 
\end{center}

\begin{center} {\it {Departamento de F\1sica,  Universidad de Burgos, 
09001 Burgos, Spain}}

e-mail: angelb@ubu.es, ablasco@ubu.es
\end{center}

  \medskip

\begin{abstract} 
\noindent

The three integrable two-dimensional H\'enon-Heiles systems and their integrable perturbations are revisited. A family of new integrable perturbations is found, and $N$-dimensional completely integrable generalizations of all these systems are constructed
by making use of $sl(2,\mathbb{R})\oplus h_3$ as their underlying Poisson symmetry algebra. In general, the procedure here introduced can be applied  in order to obtain $N$-dimensional integrable generalizations of any 2D integrable potential of the form $\mathcal{V}(q_1^2, q_2)$, and the formalism gives the explicit form of all the integrals of the motion. Further applications of this algebraic approach in different contexts are suggested.

 \end{abstract}

\bigskip\bigskip\bigskip\bigskip



\vfill
\newpage


\sect{Introduction}

The H\'enon-Heiles Hamiltonian 
\be
{H}=\dfrac{1}{2}(p_{1}^{2}+p_{2}^{2}) + \dfrac{1}{2}(q_{1}^{2}+q_{2}^{2})+\lambda\left(
q_{1}^{2}q_{2}-\frac{1}{3}\,q_{2}^{3}\right)
\label{HHaut}
\ee
was introduced in~\cite{HH} in order to model a Newtonian axially-symmetric galactic system, and it was soon considered as the paradigm of a system that exhibits chaotic behaviour (see for instance \cite{Tabor, Gutzwiller, BoPu}). Later, when its following generalization containing adjustable parameters was considered, 
\be
\mathcal{H}^{(2)}=\dfrac{1}{2}(p_{1}^{2}+p_{2}^{2})+ \delta q_{1}^{2}+(\delta+\Omega)\,q_{2}^{2}+\alpha \left(
q_{1}^{2}q_{2}+\beta\,q_{2}^{3}\right)
\label{hhmulti}
\ee
it was proven that the only Liouville--integrable members of this family of generalized 2D H\'enon-Heiles Hamiltonians
were given by the three following choices of the parameters (see~\cite{BSV}--\cite{Pickering}  and references therein):

\begin{itemize}

\item (i) The Sawada-Kotera case: $\beta=1/3$ and $\Omega=0$.

\item (ii) The KdV case: $\beta=2$ and $\Omega$ arbitrary.

\item (iii) The Kaup-Kupershmidt case: $\beta=16/3$ and $\Omega=15\,\delta$.

\end{itemize}

Here we have used the terminology of~\cite{FordyHH}, since these three integrable cases 
correspond precisely to the stationary flows of three integrable fifth-order polynomial nonlinear evolution equations. We also recall that in the KdV case the integral of the motion is quadratic in the momenta, while in the Sawada-Kotera and Kaup-Kupershmidt cases is a quartic one. Nevertheless, the separability of the latter Hamiltonian has been also been demonstrated through an involved canonical transformation in~\cite{salerno,VMC}, and the explicit integration of the three systems has been also studied (see the recent paper~\cite{KGM} and references therein). 
Therefore, H\'enon-Heiles (hereafter HH) systems are privileged benchmarks for the study of the transition between integrable and non-integrable regimes from different viewpoints, including also their quantizations~\cite{Tabor,Gutzwiller,BoPu}. As a consequence, higher dimensional generalizations of the HH systems have been considered~\cite{Hindues,KGM,2toN,RauchHH} as well as several integrable perturbations of these Hamiltonians have also been found (see, for instance~\cite{HoneIP,HonePLA} as two recent representative works for the KdV case). 

The aim of this paper is to present a novel Poisson-algebraic approach to the integrability properties of the HH systems and their perturbations, that can be straightforwardly applied to any 2D integrable potential of the form $\mathcal{V}(q_1^2, q_2)$. The cornerstone of this construction is obtained by realizing that such 2D Hamiltonians can always be defined as a specific $N=2$ symplectic realization of an `abstract' Hamitonian $\cal H$ defined on the direct-sum Poisson algebra $sl(2,\mathbb{R})\oplus h_3$. 
Afterwards, if we consider a suitable $N$-dimensional (herafter ND) symplectic realization of $\cal H$, we obtain an ND completely integrable system whose integrals of the motion will be explicitly known. This construction can be interpreted as a Poisson-algebraic generalization for the procedure given in~\cite{2toN, GrammaticosExtender} and is based on the more generic framework of the so-called coalgebra symmetry approach to integrable Hamiltonian systems (see~\cite{BR}--\cite{tesis}). In particular, the $sl(2,\mathbb{R})$ coalgebra symmetry has been recently exploited in~\cite{Annals} in order to introduce (super)integrable Kepler-Coulomb and oscillator potentials on several ND curved spaces.

The structure of the paper is as follows. In the next Section the three cases of integrable HH systems and all their known integrable generalizations are reviewed. Note that in the multiparameter framework given by (\ref{hhmulti}) the oscillator term can also be considered as a perturbation, since the oscillator constants $\delta$ and $\Omega$ are explicitly taken into account as independent parameters.
We also present a  new perturbation  (to the best of our knowledge) for the KdV case,  that can be obtained as a suitable superposition of previously known results.
In Section 3 the $sl(2,\mathbb{R})\oplus h_3$ Poisson algebra symmetry of all the 2D HH potentials is shown.  Based on this result, in Section 4 we describe the algebraic procedure in order to obtain ND integrable generalizations of the HH systems, and we show how the $sl(2,\mathbb{R})\oplus h_3$ symmetry will give us automatically the corresponding integrals of the motion.  In Section 5 the explicit expressions for the generalized ND HH systems are given, showing that new additional centrifugal/monopole terms coming from the most generic symplectic realization of $sl(2,\mathbb{R})\oplus h_3$ can be included without breaking the complete integrability of the system.
The final Section is devoted to comment on several possible applications and generalizations of the algebraic approach here presented, including the construction of quantum and curved analogues of the HH systems. We also stress that the method here presented can be easily used to get integrable ND generalizations of any 2D integrable potential of the form $\mathcal{V}(q_1^2, q_2)$, and the explicit example of the Holt potential is sketched.

\sect{Integrable perturbations of 2D H\'enon-Heiles systems}

In this Section we review all the known integrable perturbations of the 2D integrable HH systems. We will write all the cases under the same type of parameter conventions in order to provide a unified approach to the known results, which are scattered in the literature under different notations.


\subsect{The Sawada-Kotera case}

The only known perturbation of the HH-Sawada-Kotera Hamiltonian is obtained by adding a centrifugal term to the original unperturbed system, namely~\cite{Conte}
\be
\mathcal{H}^{(2)}=\dfrac{1}{2}(p_{1}^{2}+p_{2}^{2})+\delta(q_{1}^{2}+q_{2}^{2})+\alpha\,\left(q_{1}^{2}q_{2}+\dfrac{1}{3}q_{2}^{3}\right)+\dfrac{\lambda}{q_{1}^{2}}
\label{sk}
\ee
and the integral of the motion (which is quartic in the momenta) reads
\begin{eqnarray}
\mathcal{I}^{(2)}&=& \dfrac{1}{2}p_{1}^{2}p_{2}^{2}+2\delta (\delta q_{1}^{2}q_{2}^{2}+p_{1}q_{1}p_{2}q_{2})+\dfrac{2\alpha \delta q_{1}^{2}}{3}\left(
q_{1}^{2}q_{2}+3q_{2}^{3}
\right)\notag\\
&&+\alpha \left(
\alpha q_{1}^{2}q_{2}^{2}\left[
\dfrac{q_{2}^{2}}{2}+\dfrac{q_{1}^{2}}{3}
\right]+\dfrac{\alpha}{18}{q_{1}^{6}}+ p_{1}q_{1}q_{2}\left[
p_{2}q_{2}-\dfrac{2 p_{1}q_{1}}{3}
\right]+\dfrac{q_{1}^{2}}{3}\left[
2 p_{1}^{2}q_{2}+p_{1}q_{1}p_{2}\right]
\right)\notag\\
&& +\lambda \left(
\dfrac{p_{2}^{2}}{q_{1}^{2}}+\dfrac{4\alpha}{3}q_{2}
\right). \label{isk}
\end{eqnarray}
Note that we have ordered the terms in the integral with respect to the perturbation parameters. This convention will be preserved from now on, and will facilitate the immediate obtention of the corresponding integrals when some parameter(s) vanish. In particular, the integral of the motion for the unperturbed system is obtained by taking the $\lambda\to 0$ limit. Note also that the isotropic oscillator term can be easily removed by taking $\delta\to 0$.


\subsect{The Kaup-Kupershmidt case}

In this case the only integrable perturbation is given by two rational terms, namely
\be
\mathcal{H}^{(2)}=\dfrac{1}{2}\left(
p_{1}^{2}+p_{2}^{2}
\right)+\delta (q_{1}^{2}+16 q_{2}^{2})+\alpha\left(
q_{1}^{2}q_{2}+\dfrac{16}{3}q_{2}^{3}
\right)+\dfrac{\lambda}{q_{1}^{2}}+\dfrac{\nu}{q_{1}^{6}}\label{HKK}
\ee
whose constant of the motion is
\begin{eqnarray}
\mathcal{I}^{(2)}&=&\dfrac{3}{4}p_{1}^{4}+\delta (q_{1}^{2}[3 \delta q_{1}^{2}+p_{1}^{2}]+2p_{1}^{2}q_{1}^{2})\notag\\
&& +\alpha \left(
q_{1}^{2}(q_{2}p_{1}^{2}-p_{2}p_{1}q_{1})-\alpha q_{1}^{4}\left[
\dfrac{q_{1}^{2}}{6}+q_{2}^{2}
\right]+2 q_{2}(p_{1}^{2}q_{1}^{2}-\delta q_{1}^{4})
\right)\notag\\
&& +\lambda\left(
\dfrac{3}{q_{1}^{2}}\left(
p_{1}^{2}+\dfrac{\lambda}{q_{1}^{2}}
\right)+2\alpha q_{2}
\right)+\dfrac{3\nu}{q_{1}^{4}}\left(
2\alpha q_{2}+2\delta +\dfrac{1}{q_{1}^{2}}\left[
p_{1}^{2}+\dfrac{2\lambda}{q_{1}^{2}}+\dfrac{\nu}{q_{1}^{6}}
\right]
\right)\label{HKKI}
\end{eqnarray}
which is also quartic in the momenta. We recall that the $q^{-2}$ perturbation was given in \cite{2toN} while the term  $q^{-6}$ was considered in~\cite{GDRII, Hietarinta, Perelomov}. Again, the four parameters can be supressed independently and the corresponding integral of the motion would be readily obtained.


\subsect{The KdV case}

This is the integrable HH system that admits the largest family of integrable perturbations, probably due to its connections with stationary and travelling wave reductions of KdV equations \cite{FordyHH, HoneIP, HonePLA, Conte,Tondo}. Moreover, under all the perturbationes here considered the integral of the motion remains to be quadratic in the momenta.

\subsubsection{The $q^{-2}$ perturbation}

As usual, a centrifugal term can be added also in this case:
\be
\mathcal{H}^{(2)}=\dfrac{1}{2}(p_{1}^{2}+p_{2}^{2})+ \delta q_{1}^{2}+(\delta+\Omega)\,q_{2}^{2}+\alpha \left(
q_{1}^{2}q_{2}+2\,q_{2}^{3}\right)+\dfrac{\lambda}{q_{1}^{2}}
\label{kdvm2}
\ee
and the integral is
\begin{eqnarray}
\mathcal{I}^{(2)}&=&\delta \left(
\dfrac{3}{2}p_{1}^{2}+(3\delta-\Omega)q_{1}^{2}
\right)-\dfrac{\Omega}{2}p_{1}^{2}+\alpha\left(
-q_{2}p_{1}^{2}+\alpha q_{1}^{2}\left(
\dfrac{q_{1}^{2}}{4}+q_{2}^{2}
\right)+p_{2}p_{1}q_{1}
\right)\notag\\
&& +2\alpha \delta q_{2}q_{1}^{2}+\dfrac{\lambda}{q_{1}^{2}}\left(
3\delta -\Omega-2\alpha q_{2}
\right).  \label{hkdvi2}
\end{eqnarray}
Note that when all the perturbation parameters vanish, it suffices to take one of the momenta as the integral of the motion.
This system was considered in~\cite{Conte} and is also a particular case of the system II in~\cite{2toN} by taking
$ g_{4}=g_{1}=h_{3}=h_{4}=0$ and $g_{2}=\delta,4g_{2}=\delta+\Omega, g_{3}={\alpha}/{4}$. Note that the anisotropy in the oscillator term can be independently removed by taking $\Omega\to 0$.

\subsubsection{The Ramani series of polynomial deformations}

The Ramani potentials $\mathcal{V}_{M}(q_1,q_2)$ are homogeneous polynomial potentials with degree $M$ and given by~\cite{Hietarinta,RDGprl}
\be
\mathcal{V}_{i}=\sum\limits_{k=0}^{[\frac{i}{2}]}2^{i-2k}\dbinom{i-k}{k}q_{1}^{2\,k}q_{2}^{i-2k}
\qquad i=1,2,\dots
\ee
namely, the first potentials of this infinite family are given by
\bea
&& \mathcal{V}_{0}=1\\
&& \mathcal{V}_{1}=2q_{2}\\
&& \mathcal{V}_{2}=4q_{2}^2+q_{1}^2\\
&& \mathcal{V}_{3}=8q_{2}^3+4q_{1}^2q_{2}\\
&& \mathcal{V}_{4}=16 q_{2}^4+ 12 q_{1}^2 q_{2}^2 + q_{1}^4\\
&& \mathcal{V}_{5}=32q_{2}^5+32 q_{1}^2 q_{2}^3 + 6 q_{1}^4 q_{2}.
\eea
By considering 
\be
\mathcal{H}^{(2)}=\dfrac{1}{2}(p_{1}^{2}+p_{2}^{2})+\alpha_{2}\mathcal{V}_{2}+\alpha_{3}\mathcal{V}_{3}
\ee
with
$
\alpha_{2}=\delta,  \alpha_{3}={\alpha}/{4}
$
we get the KdV-HH Hamiltonian with $\Omega=3 \delta$ y $\lambda=0$. But it is well known that the full Ramani series can be superposed by preserving the complete integrability of the system (see~\cite{Hietarinta,RDGprl}) and therefore we obtain the integrable Hamiltonian given by
\bea
\mathcal{H}_M^{(2)}&=&\dfrac{1}{2}\left(
p_{1}^{2}+p_{2}^{2}
\right)+\sum\limits_{i=1}^M
\alpha_i\,\mathcal{V}_{i}\cr
&=&
\dfrac{1}{2}\left(p_{1}^{2}+p_{2}^{2}\right)+\sum\limits_{i=1}^{M}\sum\limits_{k=0}^{[\frac{i}{2}]}\alpha_{i}2^{i-2k}\dbinom{i-k}{k}q_{1}^{2k}q_{2}^{i-2k}
\eea
whose integral of the motion is:
\bea
\mathcal{I}_M^{(2)}&=&-q_{2}p_{1}^{2}+q_{1}p_{1}p_{2}
+q_{1}^{2}\sum\limits_{i=1}^{M}\alpha_{i}\mathcal{V}_{i-1} \cr
&=&-q_{2}p_{1}^{2}+q_{1}p_{1}p_{2}
+q_{1}^{2}\left(
\sum\limits_{i=1}^{M}\sum\limits_{k=0}^{[\frac{i-1}{2}]}\alpha_{i}2^{i-1-2k}\dbinom{i-1-k}{k}q_{1}^{2k}q_{2}^{i-1-2k}\right).
\eea

\subsubsect{Rational perturbations}

In the particular case $\delta=\Omega=0, \alpha={1}/{2}$ (therefore, with no oscillator potential) the following integrable perturbation of rational type is known~\cite{HonePLA} 
\bea
&& \mathcal{H}_{R}^{(2)}=\dfrac{1}{2}(p_{1}^{2}+p_{2}^{2})+\left(
\dfrac{1}{2}q_{1}^{2}q_{2}+q_{2}^{3}
\right)+2 c\, q_{2}+\sum\limits_{i=0}^{R}2^{2i+1}\xi_{i}\dfrac{\mathcal{V}_{i}}{q_{1}^{2i+2}} \cr
&& = \dfrac{1}{2}(p_{1}^{2}+p_{2}^{2})+\left(
\dfrac{1}{2}q_{1}^{2}q_{2}+q_{2}^{3}
\right)+2 c\, q_{2}+\sum\limits_{i=0}^{R}\sum\limits_{k=0}^{[\frac{i}{2}]}2^{3i-2k+1}\xi_{i}\dbinom{i-k}{k}\dfrac{q_{2}^{i-2k}}{q_{1}^{2i+2-2k}}
\eea
where $\mathcal{V}_{i}$ are the potentials in the Ramani series.
The generic integral of the motion for this system is given by
\bea
\mathcal{I}_{R}^{(2)}&=&q_{1}p_{1}p_{2}-q_{2}p_{1}^{2}+\dfrac{1}{2}q_{1}^{2}q_{2}^{2}+\dfrac{1}{8}q_{1}^{4}+c \,q_{1}^{2}-q_{1}^{2}\sum\limits_{i=0}^{R}2^{2i+1}\xi_{i}\dfrac{\mathcal{V}_{i+1}}{q_{1}^{2i+4}}
\cr
&=&q_{1}p_{1}p_{2}-q_{2}p_{1}^{2}+\dfrac{1}{2}q_{1}^{2}q_{2}^{2}+\dfrac{1}{8}q_{1}^{4}+c\, q_{1}^{2}-q_{1}^{2}\sum\limits_{i=0}^{R}\sum\limits_{k=0}^{[\frac{i+1}{2}]}2^{3i-2k+1}\xi_{i}\dbinom{i+1-k}{k}\dfrac{q_{2}^{i+1-2k}}{q_{1}^{2i+4-2k}}.\notag\\
\eea
In particular, when $R=2$ we get the integrable system
\be
\mathcal{H}^{(2)}_{2}=\dfrac{1}{2}(p_{1}^{2}+p_{2}^{2})+\left(
\dfrac{1}{2}q_{1}^{2}q_{2}+q_{2}^{3}
\right)+2 c\, q_{2}+\dfrac{2\xi_{0}}{q_{1}^{2}}+8\xi_{1}\dfrac{2 q_{2}}{q_{1}^{4}}+32\xi_{2}\dfrac{4 q_{2}^{2}+q_{1}^{2}}{q_{1}^{6}}
\ee
and for $R=3$  we recover the system (II) in~\cite{2toN} provided that 
$
g_{1}=c,  g_{2}=0,  g_{3}={1}/{8},  2\xi_{0}=\lambda,  h_{3}=8\xi_{1},  h_{4}=32 \xi_{2}
$.



\subsubsect{New integrable perturbations}

But it turns out that all the two latter perturbations of the KdV case can be superposed. This novel result can be immediately checked by direct computation, and stresses again the ductility of the KdV-HH system, that can be perturbed in different directions by preserving its integrability.
In particular, the rational perturbations can also be included in the full Ramani series provided that $M>R$. This encompasses all the previous KdV perturbations with $\Omega=3 \delta$ and leads to the new system:
\begin{eqnarray}
\mathcal{H}_{M, R}^{(2)}&=&\dfrac{1}{2}\left(p_{1}^{2}+p_{2}^{2}\right)+\dfrac{\lambda}{q_{1}^{2}}+\sum\limits_{i=1}^{M}\alpha_{i}\mathcal{V}_{i}+\sum\limits_{i=1}^{R}\gamma_{i}\dfrac{\mathcal{V}_{i}}{q_{1}^{2i+2}}
\cr
&=&\dfrac{1}{2}\left(p_{1}^{2}+p_{2}^{2}\right)+\dfrac{\lambda}{q_{1}^{2}}+\sum\limits_{i=1}^{M}\sum\limits_{k=0}^{[\frac{i}{2}]}\alpha_{i}2^{i-2k}\dbinom{i-k}{k}q_{1}^{2k}q_{2}^{i-2k}\notag\\
&& +\sum\limits_{i=1}^{R}\sum\limits_{k=0}^{[\frac{i}{2}]}\gamma_{i}2^{i-2k}\dbinom{i-k}{k}\dfrac{q_{2}^{i-2k}}{q_{1}^{2i+2-2k}}\,\,\,\,\, (M>R).\label{KdVFP}
\end{eqnarray}
The corresponding integral of the motion can also be obtained by direct computation and reads
\begin{eqnarray}
\mathcal{I}_{M, R}^{(2)}&=&-q_{2}p_{1}^{2}+q_{1}p_{1}p_{2}-\dfrac{2\lambda}{q_{1}^{2}}q_{2}+q_{1}^{2}\left(
\sum\limits_{i=1}^{M}\alpha_{i}\mathcal{V}_{i-1}-\sum\limits_{i=1}^{R}\gamma_{i}\dfrac{\mathcal{V}_{i+1}}{q_{1}^{2i+4}}
\right)\,\,\,\,
\cr
&=&-q_{2}p_{1}^{2}+q_{1}p_{1}p_{2}-\dfrac{2\lambda}{q_{1}^{2}}q_{2}\notag\\
&& +q_{1}^{2}\left(
\sum\limits_{i=1}^{M}\sum\limits_{k=0}^{[\frac{i-1}{2}]}\alpha_{i}2^{i-1-2k}\dbinom{i-1-k}{k}q_{1}^{2k}q_{2}^{i-1-2k}\right)\notag\\
&& -q_{1}^{2}\left(\sum\limits_{i=1}^{R}\sum\limits_{k=0}^{[\frac{i+1}{2}]}\gamma_{i}2^{i+1-2k}\dbinom{i+1-k}{k}\dfrac{q_{2}^{i+1-2k}}{q_{1}^{2i+4-2k}}
\right)\,\,\,(M>R).\label{KdVFPI}
\end{eqnarray}
As a particular example, if we consider the case $M=4, R=3$ we obtain
\begin{eqnarray}
\mathcal{H}^{(2)}_{4, 3}&=&\dfrac{1}{2}(p_{1}^{2}+p_{2}^{2})+\alpha_{1}(2q_{2})+\alpha_{2}(4q_{2}^2+q_{1}^2)+4\alpha_{3}(2q_{2}^3+q_{1}^2q_{2})\notag\\
&&+\alpha_{4}(16 q_{2}^4+ 12 q_{1}^2 q_{2}^2 + q_{1}^4) +\dfrac{\lambda}{q_{1}^{2}}+\gamma_{1}\dfrac{2q_{2}}{q_{1}^{4}}+\gamma_{2}\dfrac{(4q_{2}^2+q_{1}^2)}{q_{1}^{6}}+\gamma_{3}\dfrac{(8q_{2}^3+4q_{1}^2q_{2})}{q_{1}^{8}}\notag\\
\end{eqnarray}
where the notation in~\cite{HonePLA} corresponds to $
\lambda=2 \xi_{0},
\gamma_{j}=2^{2j+1}\xi_{j}\,\,\,\,(j=1,2,\ldots)$.
As it can be easily appreciated, $M$ is just the degree of the polynomial perturbation and $(2 R+2)$ is the maximum degree in the denominator of the rational perturbations.


\sect{Hamiltonian systems on the Poisson algebra $sl(2,\mathbb{R})\oplus h_3$}

The aim of this Section is to show that all the abovementioned integrable systems can be written within a common algebraic framework as Hamiltonian systems on the six-dimensional Poisson algebra $sl(2,\mathbb{R})\oplus h_3$, with generators $\{J_{+},J_{-},J_{3},A_+,A_-,M\}$ and non-vanishing Poisson brackets given by
\begin{equation}
 \{J_3,J_+\}=2 J_+     \qquad 
\{J_3,J_-\}=-2 J_- \qquad   
\{J_-,J_+\}=4 J_3    \qquad \{A_-,A_+\}=M
\label{bafinal}
\end{equation}
(all the brackets between generators of $sl(2,\mathbb{R})$ and $h_{3}$ vanish). Obviously, this Poisson algebra has two Casimir functions: the central generator $M$ and the $sl(2,\mathbb{R})$ Casimir function
\be
\mathcal{C}_{sl(2)}=J_{+}J_{-}-J_{3}^{2}.
\ee
As we shall demonstrate in the next Section, this underlying symmetry of the integrable 2D HH systems will provide ND generalizations for all of them in a systematic way.

In particular, let us consider the two-particle symplectic realization of $sl(2,\mathbb{R})\oplus h_3$ given by
\be
 J_{+}=p_{1}^{2}\qquad
 J_{-}=q_{1}^{2}\qquad
 J_{3}=q_{1}p_{1}
\label{ssl22}
\ee
\be 
A_+=  p_2 \qquad
A_-=q_2  \qquad   M=1.
\label{sh62}
\ee 
whose symplectic leaves will be labelled by the values of the Casimir functions, namely $\mathcal{C}_{sl(2)}=0$ and $M=1$. Now it is easy to realize that all the 2D integrable HH Hamiltonians previously described are of the form
\be
\mathcal{H}^{(2)}=\dfrac{1}{2}(p_{1}^2+p_{2}^2)+\mathcal{V}(q_1^2, q_2).
\label{hxy}
\ee
Now it becomes clear that $\mathcal{H}^{(2)}$ is $sl(2,\mathbb{R})\oplus h_3$ symmetric, since it would be just the $N=2$ realization (\ref{ssl22}) and (\ref{sh62}) of the generic Hamiltonian
\be
\mathcal{H}=\dfrac{1}{2}(J_{+}+A_{+}^2)+\mathcal{V}(J_{-}, A_{-}).
\label{hsl6h6}
\ee
Moreover, for each specific choice of $\mathcal{H}$ that corresponds to a given integrable HH system, it turns out that we can find an  ``abstract" function ${\cal I}={\cal I}(J_{+},J_{-},J_{3},A_+,A_-,M )$ in involution with $\mathcal{H}$ with respect to the Poisson bracket (\ref{bafinal}) and whose symplectic realization (\ref{ssl22})--(\ref{sh62}) is just the associated integral of the motion $\mathcal{I}^{(2)}$. In fact, 
the only thing that we have to do is to write the corresponding $\mathcal{I}^{(2)}$ as the most generic function of the form
\be
\mathcal{I}^{(2)}=\mathcal{I}^{(2)}(p_{1}^2,\ q_1^2,\ q_1\,p_{1}, \ q_2, \,p_{2})
\label{ixy}
\ee
and perform the substitution
\be
\mathcal{I}^{(2)}(p_{1}^2,\ q_1^2,\ q_1\,p_{1}, \ q_2, \,p_{2})\rightarrow
{\cal I}(J_{+},\ J_{-},\ J_{3},\ A_+,\ A_- ).\label{ixy}
\ee
Here by `most generic' substitution we mean that, for instance, terms of the type $q_{1}^2\,p_{1}^2$ have to be replaced by a linear combination of the type $a_1\,J_{+}\,J_{-} + a_2\,J_{3}^2$.
Finally one has to impose that the function ${\cal I}$ so obtained does Poisson commutes with (\ref{hsl6h6}) by using the bracket (\ref{bafinal}), and this process will fix unambiguously all the constants $a_i$. In this way the  `abstract'  integral of the motion $\mathcal{I}$ is obtained, thus obtaining explicitly the full $sl(2,\mathbb{R})\oplus h_3$ invariance of the system. As we shall see in the sequel, this scheme can be sucessfully applied onto all the integrable HH systems that have been previously described.


\subsect{The Sawada-Kotera case}

If we consider the $N=2$ realization (\ref{ssl22}) and (\ref{sh62}) of the algebra $sl(2,\mathbb{R})\oplus h_3$ we can write the perturbed Sawada-Kotera Hamiltonian (\ref{sk}) as
\be
\mathcal{H}=\dfrac{1}{2}\left(
J_{+}+A_{+}^{2}
\right)+\delta (J_{-}+ A_{-}^{2})+\alpha \left(
J_{-}A_{-}+\dfrac{1}{3}A_{-}^{3}
\right)+\dfrac{\lambda}{J_{-}}
\label{hska}
\ee
and the integral of the motion (\ref{isk}) can be writen as an `abstract' $sl(2,\mathbb{R})\oplus h_3$ funtion in the following way:
\begin{eqnarray}
\mathcal{I}&=&\dfrac{1}{2}J_{+}A_{+}^{2}+2\delta A_{-}(\delta A_{-}J_{-}+A_{+}J_{3})+2\alpha \delta A_{-}J_{-}\left(
\dfrac{J_{-}}{3}+A_{-}^{2}
\right)\notag\\
&& +\alpha\left(
\alpha A_{-}^{2}J_{-}\left[\dfrac{A_{-}^{2}}{2}+\dfrac{J_{-}}{3}\right]+\dfrac{\alpha }{18}J_{-}^{3}+A_{-}J_{3}\left[
A_{+}A_{-}-\dfrac{2J_{3}}{3}
\right]+\dfrac{J_{-}}{3}\left[
2 A_{-}J_{+}+A_{+}J_{3}
\right]
\right)\notag\\
&& +\lambda \left(
\dfrac{A_{+}^{2}}{J_{-}}+\dfrac{4\alpha}{3}A_{-}
\right).
\label{hski}
\end{eqnarray}
By making use of the Poisson algebra (\ref{bafinal}), a straightforward computation shows that (\ref{hska}) and (\ref{hski}) are in involution, and this will be true for any further symplectic realization of the $sl(2,\mathbb{R})\oplus h_{3}$ Poisson algebra that we could consider.

\subsect{The Kaup-Kupershmidt case}

In this case the very same procedure for (\ref{HKK}) gives us the $sl(2,\mathbb{R})\oplus h_3$ invariant object:
\be
\mathcal{H}=\dfrac{1}{2}\left(
J_{+}+A_{+}^{2}
\right)+\delta (J_{-}+16 A_{-}^{2})+\alpha\left(
J_{-}A_{-}+\dfrac{16}{3}A_{-}^{3}
\right)+\dfrac{\lambda}{J_{-}}+\dfrac{\nu}{J_{-}^{3}}\label{HKKa}
\ee
and the integral of the motion coming from (\ref{HKKI}) is proven to be
\begin{eqnarray}
\mathcal{I}&=& \dfrac{3}{4}J_{+}^{2}+\delta (J_{-}[3\delta J_{-}+J_{+}]+2 J_{3}^{2})\notag\\
&& +\alpha \left(
J_{-}(A_{-}J_{+}-A_{+}J_{3})-\alpha J_{-}^{2}\left[
\dfrac{J_{-}}{6}+A_{-}^{2}
\right]+2 A_{-}(J_{3}^{2}-\delta J_{-}^{2})
\right)\notag\\
&& +\lambda \left(
\dfrac{3}{J_{-}}\left(
J_{+}+\dfrac{\lambda}{J_{-}}
\right)+2\alpha A_{-}
\right)+\dfrac{3\nu}{J_{-}^{2}}\left(2\alpha A_{-}+2\delta +\dfrac{1}{J_{-}}\left[
J_{+}+\dfrac{2\lambda}{J_{-}}+\dfrac{\nu}{J_{-}^{3}}
\right]
\right). \label{HKKIa}
\notag\\ 
\end{eqnarray}
Once again, the involutivity between (\ref{HKKIa}) and (\ref{HKKa}) can be easily checked.

\subsect{The KdV case}

In this case the HH Hamiltonian (\ref{kdvm2}) is given by
\be
\mathcal{H}=\dfrac{1}{2}(J_{+}+A_{+}^{2})+\delta (J_{-}+A_{-}^{2})+\Omega A_{-}^{2}+\alpha (
J_{-}A_{-}+2 A_{-}^{3}
)+\dfrac{\lambda}{J_{-}}
\label{hkdv}
\ee 
and (\ref{hkdvi2}) is just the symplectic realization of the `abstract' integral
\begin{eqnarray}
\mathcal{I}&=&\delta \left(\dfrac{3}{2}J_{+}+(3 \delta -\Omega)J_{-}\right)-\dfrac{\Omega}{2}J_{+}+\alpha\left(-A_{-}J_{+}+\alpha J_{-}\left(
\dfrac{J_{-}}{4}+A_{-}^{2}
\right)+
A_{+}J_{3}
\right)\notag\\
&&+ 2 \alpha \delta A_{-}J_{-}+\dfrac{\lambda}{J_{-}}(3\delta -\Omega -2 \alpha A_{-}).
\label{hkdvi}
\end{eqnarray}
In the same way, the full perturbed KdV system (\ref{KdVFP}) is written in terms of the Poisson algebra generators as
\bea
\mathcal{H}_{M, R}&=&\dfrac{1}{2}\left(J_{+}+A_{+}^{2}\right)+\dfrac{\lambda}{J_{-}}+\sum\limits_{i=1}^{M}\alpha_{i}\mathcal{V}_{i}+\sum\limits_{i=1}^{R}\dfrac{\gamma_{i}}{J_{-}^{i+1}}\mathcal{V}_{i}\,\cr
&=&\dfrac{1}{2}\left(J_{+}+A_{+}^{2}\right)+\dfrac{\lambda}{J_{-}}+\sum\limits_{i=1}^{M}\sum\limits_{k=0}^{[\frac{i}{2}]}\alpha_{i}2^{i-2k}\dbinom{i-k}{k}J_{-}^{k}A_{-}^{i-2k}\notag\\
&& +\sum\limits_{i=1}^{R}\sum\limits_{k=0}^{[\frac{i}{2}]}\gamma_{i}2^{i-2k}\dbinom{i-k}{k}\dfrac{A_{-}^{i-2k}}{J_{-}^{i+1-k}}\,\,\,\,\, (M>R)
\label{HKvDFPa}
\eea
and the integral of the motion coming from (\ref{KdVFPI}) is found to be
\begin{eqnarray}
\mathcal{I}_{M, R}&=&-A_{-}J_{+}+J_{3}A_{+}-\dfrac{2\lambda}{J_{-}}A_{-}\notag\\
&& +J_{-}\left(
\sum\limits_{i=1}^{M}\sum\limits_{k=0}^{[\frac{i-1}{2}]}\alpha_{i}2^{i-1-2k}\dbinom{i-1-k}{k}J_{-}^{k}A_{-}^{i-1-2k}\right)\notag\\
&& -J_{-}\left(\sum\limits_{i=1}^{R}\sum\limits_{k=0}^{[\frac{i+1}{2}]}\gamma_{i}2^{i+1-2k}\dbinom{i+1-k}{k}\dfrac{A_{-}^{i+1-2k}}{J_{-}^{i+2-k}}
\right)\,\,\,(M>R). 
\label{HKvDFPIa}
\end{eqnarray}


\sect{$N$-dimensional symplectic realizations}

The previous $sl(2,\mathbb{R})\oplus h_3$ Hamiltonian structure of the HH systems will allow us to construct ND integrable generalizations of them in a systematic way. In particular, we can firstly introduce the following ND symplectic realization of $sl(2,\mathbb{R})\oplus h_3$, in which the first set of $(N-1)$ degrees of freedom is associated to $sl(2,\mathbb{R})$, while the last one is just the usual one-particle $h_3$ realization with $M=1$:
\be
 J_{+}=\sum\limits_{i=1}^{N-1}
p_{i}^{2}\qquad
 J_{-}=\sum\limits_{i=1}^{N-1}q_{i}^{2}\qquad
 J_{3}=\sum\limits_{i=1}^{N-1}q_{i}p_{i}
\label{ssl2}
\ee
\be 
A_+=  p_N \qquad
A_-=q_N  \qquad   M=1.
\label{sh6}
\ee 
It is straightforward to prove that any ND Hamiltonian constructed as the previous symplectic realization of any function $\cal H$ on $sl(2,\mathbb{R})\oplus h_3$, namely
\begin{equation}
{\cal H}^{(N)}={\cal H}(J_{+},J_{-},J_{3} ,A_+ ,A_-,M )
\label{hslh6}
\end{equation}
is {\em quasi-integrable}, since the following $(N-2)$ functions 
\be
\mathcal{C}^{(m)}=\sum\limits_{1\leq i <j}^{m}\left(
q_{i}p_{j}-q_{j}p_{i}
\right)^{2}\,\,\,\,\,\, m=2,\ldots, N-1
\label{sl2cas}
\ee
are constants of the motion for ${\cal H}$, and they are mutually in involution. In fact (\ref{sl2cas}) are just the so-called $sl(2,\mathbb{R})$-coalgebra integrals for the $sl(2,\mathbb{R})$-sector of the Hamiltonian (see~\cite{BR,  CRMAngel, alfonso, RutwigProc} for details). Note that all these constants of the motion depend on the first $(N-1)$ canonical variables, and any contribution in the Hamiltonian coming from the last degree of freedom $(q_N,p_N)$ will obviously Poisson-commute with any of them.

Therefore, any system given by (\ref{hslh6}) is only `one integral away' from being completely integrable. Moreover, if  for a certain choice of ${\cal H}$ we are able to find one additional independent integral of the `abstract' form 
\begin{equation}
{\cal I}={\cal I}(J_{+},J_{-},J_{3},A_+,A_-,M )
\label{invslh6}
\end{equation}
and we realize it under the ND symplectic realization (\ref{ssl2}) and (\ref{sh6}), the complete integrability of ${\cal H}^{(N)}$ will be guaranteed in the ND case.
This will be the case for all the HH systems, whose 2D integrability will provide such function $\cal I$.

Thus, the ND analogue of a given 2D HH system will be defined as the $N$-degrees of freedom realization (\ref{ssl2})-(\ref{sh6}) of the corresponding $sl(2,\mathbb{R}) \oplus h_{3}$ Hamiltonian (\ref{hsl6h6}), and the $N$-particle realization of the additional integral ${\cal I}$ will provide the complete integrability of the system.


\subsect{A KdV example}

In order to illustrate this general result, let us consider (\ref{hkdv}) as the algebraic definition of a ND HH-KdV Hamiltonian. 
Namely, by taking $N=3$ in (\ref{ssl2})--(\ref{sh6}) and, afterwards, by substituting these expressions into (\ref{hkdv}) we get:
\be
\mathcal{H}^{(N)}=\dfrac{1}{2}\left(\sum\limits_{j=1}^{2}p_{j}^{2}+p_{3}^{2}\right)+\delta \left(\sum\limits_{j=1}^{2}q_{j}^{2}+q_{3}^{2}\right)+\Omega q_{3}^{2} +\alpha \left(
\left[\sum\limits_{j=1}^{2}q_{j}^{2}\right]q_{3}+2 q_{3}^{3}
\right)+\dfrac{\lambda}{\left(\sum\limits_{j=1}^{2}q_{j}^{2}\right)}.
\label{hkdv3D}
\ee
Note that in this way we are obtaining a `radial' generalization of the KdV potential in the $(q_1,q_2)$ subspace, which is a direct consequence of the spherical symmetry of the symplectic realization (\ref{ssl2}) for the $sl(2,\mathbb{R})$ sector.

In general, the specialization of (\ref{hkdv}) and (\ref{hkdvi}) through the $N$-particle symplectic realization (\ref{ssl2})--(\ref{sh6}) will provide the explicit form of an integrable ND generalization of the KdV Hamiltonian, that reads
\be
\mathcal{H}^{(N)}=\dfrac{1}{2}\left(\sum\limits_{j=1}^{N-1}p_{j}^{2}+p_{N}^{2}\right)+\delta \left(\sum\limits_{j=1}^{N-1}q_{j}^{2}+q_{N}^{2}\right)+\Omega q_{N}^{2}+\alpha \left(
\left[\sum\limits_{j=1}^{N-1}q_{j}^{2}\right]q_{N}+2 q_{N}^{3}
\right)+\dfrac{\lambda}{\left(\sum\limits_{j=1}^{N-1}q_{j}^{2}\right)}
\label{hkdvND}
\ee
and the explicit form of the integral of the motion coming from (\ref{hkdvi}) is immediately obtained:
\begin{eqnarray}
\mathcal{I}^{(N)}\!\!\!\!&=&\!\!\!\! \delta \left(\dfrac{3}{2}\sum\limits_{j=1}^{N-1}p_{j}^{2}+(3 \delta -\Omega)\left[\sum\limits_{j=1}^{N-1}q_{j}^{2}\right]\right)-\dfrac{\Omega}{2}\sum\limits_{j=1}^{N-1}p_{j}^{2}\notag\\
&&\!\! +\alpha\! \left(\!\!-q_{N}\sum\limits_{j=1}^{N-1}p_{j}^{2}+\alpha \left(\sum\limits_{j=1}^{N-1}q_{j}^{2}\right)\! \!\left(
\dfrac{\sum\limits_{j=1}^{N-1}q_{j}^{2}}{4}+q_{N}^{2}\!
\right)\!\!+\!\!
p_{N}\left(\sum\limits_{j=1}^{N-1}p_{j}q_{j}\right)\!\!
\right)\notag\\
&&+ 2 \alpha \delta q_{N}\left(
\sum\limits_{j=1}^{N-1}q_{j}^{2}
\right)+\dfrac{\lambda}{\left(
\sum\limits_{j=1}^{N-1}q_{j}^{2}
\right)}(3\delta -\Omega -2 \alpha q_{N}).
\label{hkdviND}
\end{eqnarray}
It is straightforward to check that (\ref{hkdviND}) Poisson-commutes with the full set of $sl(2,\mathbb{R})$ integrals (\ref{sl2cas}) and, obviously, with (\ref{hkdvND}).

\subsect{Adding centrifugal terms}

But a further generalization of this construction is still possible by considering a more general symplectic realization of the Poisson algebra $sl(2,\mathbb{R})\oplus h_3$.  Such generalization breaks the radial symmetry in the $sl(2,\mathbb{R})$ sector by adding centrifugal terms, albeit by preserving the complete integrability of the new system (note that, in general, this is by no means guaranteed when centrifugal/monopole terms are added~\cite{KCJPA}). Namely, let us consider the functions
\be
 J_{+}=\sum\limits_{i=1}^{N-1}\left(
p_{i}^{2}+\dfrac{\otra_{i}}{q_{i}^{2}}\right)\qquad
 J_{-}=\sum\limits_{i=1}^{N-1}q_{i}^{2}\qquad
 J_{3}=\sum\limits_{i=1}^{N-1}q_{i}p_{i}
 \label{sympbi}
\ee
\be 
A_+=  p_N \qquad
A_-=q_N  \qquad   M=1.
\label{sh6bi}
\ee 
which define the $(\otra_1,\otra_2,\dots,\otra_{N-1})\oplus (1)$ symplectic realization of $sl(2,\mathbb{R})\oplus h_3$. Note that its only difference with respect to (\ref{ssl2}) and (\ref{sh6}) is the addition of $(N-1)$ centrifugal terms in the realization of the $J_{+}$ generator (despite the fact that the constants $\otra_i$ can be negative, we shall continue calling those terms as `centrifugal' ones).

When these terms are included, the constants of the motion (\ref{sl2cas}) coming from the  $sl(2,\mathbb{R})$-sector of the Hamiltonian are converted into~\cite{CRMAngel, RutwigProc, Annals}
\be
\mathcal{C}^{(m)}=\sum\limits_{1\leq i <j}^{m}\left(
q_{i}p_{j}-q_{j}p_{i}
\right)^{2}+\sum\limits_{1\leq i<j}^{m}\left(\otra_{i}\dfrac{q_{j}^{2}}{q_{i}^{2}}+\otra_{j}\dfrac{q_{i}^{2}}{q_{j}^{2}}\right)\,\,\,\,\,\, m=2,\ldots, N-1.
\label{sl2casbi}
\ee
But it is essential to stress that any $sl(2,\mathbb{R})\oplus h_3$ Hamiltonian (\ref{hslh6}), when written in terms of the symplectic realization (\ref{sympbi})-(\ref{sh6bi}) will Poisson-commute with all the integrals (\ref{sl2casbi}). And the same will happen to the corresponding symplectic realization of the additional integral $\mathcal{I}$, that will be modified only through the centrifugal terms coming from the $J_{+}$ generator.

Namely, if we consider the $N=3$ generalized symplectic realization (\ref{sympbi})--(\ref{sh6bi}) of the KdV Hamiltonian (\ref{hkdv}),  we obtain
\be
\mathcal{H}^{(N)}=\dfrac{1}{2}\sum\limits_{j=1}^{2}\left( p_{j}^{2}+\frac{b_j}{q_{j}^{2}}\right)+\dfrac{1}{2}p_{3}^{2}+ \delta \left(\sum\limits_{j=1}^{2}q_{j}^{2}+q_{3}^{2}\right)+\Omega q_{3}^{2} +\alpha \left(
\left[\sum\limits_{j=1}^{2}q_{j}^{2}\right]q_{3}+2 q_{3}^{3}
\right)+\dfrac{\lambda}{\left(\sum\limits_{j=1}^{2}q_{j}^{2}\right)}.
\label{hkdvb1b2}
\ee
This system is, by construction, completely integrable: its integrals of the motion in involution are the $\mathcal{C}^{(2)}$ integral given by (\ref{sl2casbi}) and the  $N=3$ generalized symplectic realization (\ref{sympbi})--(\ref{sh6bi}) of the `abstract' integral (\ref{hkdvi}). Note that this Hamiltonian has {\em two} types of `centrifugal' terms, one of them linked to the parameter $\lambda$ and the other one induced by the generalized symplectic realization and controlled by the parameters $b_i$. We recall that when $\lambda=0$ the ND generalization of (\ref{hkdvb1b2}) has been shown to be integrable even when the harmonic oscillator term is fully anisotropic (see~\cite{KGM} and references therein). However, in our case the $sl(2,\mathbb{R})\oplus h_3$ symmetry imposes the isotropy in the first $(N-1)$ coordinates, thus allowing the existence of the `global' centrifugal term linked to $\lambda$.


\sect{ND integrable H\'enon-Heiles systems}

Finally, we present the complete expressions for the most general HH systems that we can obtain through the algebraic procedure described in the previous Sections. We stress that the following three ND Hamiltonians do commute with the {\em same} `universal' set of integrals (\ref{sl2casbi}). 
Also, it becomes evident that finding in each case the remaining integral $\mathcal{I}^{(N)}$ without the aid of the underlying $sl(2,\mathbb{R})\oplus h_3$ symmetry would be a quite complicated task. On the other hand, this symmetry guarantees by construction that the integral $\mathcal{I}^{(N)}$ is in involution with the set of functions (\ref{sl2casbi}).

\subsect{The generalized ND system of Sawada-Kotera type}

When the symplectic realization (\ref{sympbi})--(\ref{sh6bi}) is applied onto (\ref{hska}), the ND HH-Sawada-Kotera Hamiltonian reads
\be
\mathcal{H}^{(N)}=\dfrac{1}{2}\left(\sum\limits_{i=1}^{N-1}\left[p_{i}^{2}+\dfrac{b_{i}}{q_{i}^{2}}\right]+p_{N}^{2}\right)+\delta\left(
\sum\limits_{i=1}^{N-1}q_{i}^{2}+q_{N}^{2}
\right)+\alpha\left(
q_{N}\left[\sum\limits_{i=1}^{N-1}q_{i}^{2}
\right]+\dfrac{1}{3}q_{N}^{3}
\right)+\dfrac{\lambda}{\left(\sum\limits_{i=1}^{N-1}q_{i}^{2}\right)}
\ee
and the integral of the motion coming from (\ref{hski}) now reads:
\begin{eqnarray}
\mathcal{I}^{(N)}&=&\dfrac{1}{2}p_{N}^{2}
\sum\limits_{i=1}^{N-1}\left(p_{i}^{2}+\dfrac{b_{i}}{q_{i}^{2}}
\right)+2\delta q_{N}\left(
\delta q_{N}\sum\limits_{i=1}^{N-1}q_{i}^{2}+p_{N}\sum\limits_{i=1}^{N-1}q_{i}p_{i}
\right)
\notag\\
&& + 2\alpha \delta q_{N}\left[\sum\limits_{i=1}^{N-1}q_{i}^{2}\right]\left(
\dfrac{1}{3}\sum\limits_{i=1}^{N-1}q_{i}^{2}+q_{N}^{2}
\right)\notag\\
&& +\alpha\left(
\alpha\, q_{N}^{2}\left(\sum\limits_{i=1}^{N-1}q_{i}^{2}\right)\left[
\dfrac{q_{N}^{2}}{2}+\dfrac{1}{3}\sum\limits_{i=1}^{N-1}q_{i}^{2}
\right]+\dfrac{\alpha}{18}\left(\sum\limits_{i=1}^{N-1}q_{i}^{2}\right)^{3}\right)\notag\\
&& + \alpha\, q_{N}\left(\sum\limits_{i=1}^{N-1}q_{i}p_{i}\right)\left[
q_{N}p_{N}-\dfrac{2}{3}\sum\limits_{i=1}^{N-1}q_{i}p_{i}
\right]\notag\\
&& + \dfrac{ \alpha}{3}\left(\sum\limits_{i=1}^{N-1}q_{i}^{2}\right)\left[
2q_{N}\sum\limits_{i=1}^{N-1}\left(p_{i}^{2}+\dfrac{b_{i}}{q_{i}^{2}}\right)+p_{N}\left(
\sum\limits_{i=1}^{N-1}q_{i}p_{i}
\right)
\right]
\notag\\
&&+\lambda\left(
\dfrac{p_{N}^{2}}{\sum\limits_{i=1}^{N-1}q_{i}^{2}}+\dfrac{4\alpha}{3}q_{N}
\right) .
\end{eqnarray}

The fact that both quantities are in involution can be straightforwardly checked by direct computation.


\subsect{The generalized ND system of Kaup-Kupershmidt type}

In the same way we can obtain the generalized version of this system by considering the symplectic realization (\ref{sympbi})--(\ref{sh6bi}) of the `abstract' Kaup-Kupershmidt Hamiltonian (\ref{HKKa})
\begin{eqnarray}
\mathcal{H}^{(N)}&=&\dfrac{1}{2}\left(\sum\limits_{i=1}^{N-1}\left[p_{i}^{2}+\dfrac{b_{i}}{q_{i}^{2}}\right]+p_{N}^{2}\right)+\delta\left(
\sum\limits_{i=1}^{N-1}q_{i}^{2}+16q_{N}^{2}
\right)\notag\\
&&+\alpha\left(
q_{N}\left[\sum\limits_{i=1}^{N-1}q_{i}^{2}
\right]+\dfrac{16}{3}q_{N}^{3}
\right)+\dfrac{\lambda}{\left(\sum\limits_{i=1}^{N-1}q_{i}^{2}\right)}+\dfrac{\nu}{\left(\sum\limits_{i=1}^{N-1}q_{i}^{2}\right)^{3}}.
\end{eqnarray}
In this case the integral of the motion (\ref{HKKIa}) reads
\begin{eqnarray}
\mathcal{I}^{(N)}&=&\dfrac{3}{4}\left(
 \sum\limits_{i=1}^{N-1}\left[p_{i}^{2}+\dfrac{b_{i}}{q_{i}^{2}}\right]
\right)^{2}+\delta \left(
 \sum\limits_{i=1}^{N-1}q_{i}^{2}\left[
 3\delta  \sum\limits_{i=1}^{N-1}q_{i}^{2}+ \sum\limits_{i=1}^{N-1}\left(p_{i}^{2}+\dfrac{b_{i}}{q_{i}^{2}}\right)
 \right]+2\left( \sum\limits_{i=1}^{N-1}q_{i}p_{i}\right)^{2}
\right)\notag\\
&& +\alpha\left(
 \sum\limits_{i=1}^{N-1}q_{i}^{2}\left(
 q_{N} \sum\limits_{i=1}^{N-1}\left(p_{i}^{2}+\dfrac{b_{i}}{q_{i}^{2}}\right)-p_{N} \sum\limits_{i=1}^{N-1}q_{i}p_{i}
 \right)-\alpha \left( \sum\limits_{i=1}^{N-1}q_{i}^{2}\right)^{2}\left[
 \dfrac{1}{6} \sum\limits_{i=1}^{N-1}q_{i}^{2}+q_{N}^{2}
 \right]
\right)\notag\\
&&+2\,\alpha\, q_{N}\left[
\left( \sum\limits_{i=1}^{N-1}q_{i}p_{i}\right)^{2}-\delta \left( \sum\limits_{i=1}^{N-1}q_{i}^{2}\right)^{2}
\right]\notag\\
&&+\lambda\left(
\dfrac{3}{ \left(\sum\limits_{i=1}^{N-1}q_{i}^{2}\right)}\left[
 \sum\limits_{i=1}^{N-1}\left(p_{i}^{2}+\dfrac{b_{i}}{q_{i}^{2}}\right)+\dfrac{\lambda}{\left( \sum\limits_{i=1}^{N-1}q_{i}^{2}\right)}
\right]+2 \alpha q_{N}
\right)
\notag\\
&&+\dfrac{3\nu}{\left( \sum\limits_{i=1}^{N-1}q_{i}^{2}\right)^{2}}\left(
2\alpha q_{N}+2\delta +\dfrac{1}{\left( \sum\limits_{i=1}^{N-1}q_{i}^{2}\right)}\left[
 \sum\limits_{i=1}^{N-1}\left(p_{i}^{2}+\dfrac{b_{i}}{q_{i}^{2}}\right)+\dfrac{2\lambda}{\left( \sum\limits_{i=1}^{N-1}q_{i}^{2}\right)}+\dfrac{\nu}{\left( \sum\limits_{i=1}^{N-1}q_{i}^{2}\right)^{3}}
\right]
\right).\notag\\
\end{eqnarray}


\subsect{The generalized ND KdV system}

Finally, the generalized version of the $\mathcal{H}_{M, R}^{(N)}$ Hamiltonian is obtained by substituting the generalized symplectic realization into (\ref{HKvDFPa}). The final expression is
\begin{eqnarray}
\mathcal{H}_{M, R}^{(N)}&=&\dfrac{1}{2}\left(\sum\limits_{j=1}^{N-1}\left[p_{j}^{2}+\dfrac{b_{j}}{q_{j}^{2}}\right]+p_{N}^{2}\right)+\dfrac{\lambda}{\left(\sum\limits_{j=1}^{N-1}q_{j}^{2}\right)}\notag\\
&& +\sum\limits_{i=1}^{M}\sum\limits_{k=0}^{[\frac{i}{2}]}\alpha_{i}2^{i-2k}\dbinom{i-k}{k}\left(\sum\limits_{j=1}^{N-1}q_{j}^{2}\right)^{k}q_{N}^{i-2k}\notag\\
&& +\sum\limits_{i=1}^{R}\sum\limits_{k=0}^{[\frac{i}{2}]}\gamma_{i}2^{i-2k}\dbinom{i-k}{k}\dfrac{q_{N}^{i-2k}}{\left(\sum\limits_{j=1}^{N-1}q_{j}^{2}\right)^{i+1-k}}\,\,\,\,\, (M>R)
\end{eqnarray}
and the same procedure for (\ref{HKvDFPIa}) gives rise to the additional integral
\begin{eqnarray}
\mathcal{I}_{M, R}^{(N)}\!\!\!\!&=\!\!\!\!&-q_{N}\sum\limits_{j=1}^{N-1}\left(p_{j}^{2}+\dfrac{b_{j}}{q_{j}^{2}}\right)+p_{N}\left(\sum\limits_{j=1}^{N-1}p_{j}q_{j}\right)-\dfrac{2\lambda}{\left(\sum\limits_{j=1}^{N-1}q_{j}^{2}\right)}q_{N}\notag\\
&& +\left(\sum\limits_{j=1}^{N-1}q_{j}^{2}\right)\left(
\sum\limits_{i=1}^{M}\sum\limits_{k=0}^{[\frac{i-1}{2}]}\alpha_{i}2^{i-1-2k}\dbinom{i-1-k}{k}\left(\sum\limits_{j=1}^{N-1}q_{j}^{2}\right)^{k}q_{N}^{i-1-2k}\right)\notag\\
&& -\left(\sum\limits_{j=1}^{N-1}q_{j}^{2}\right)\!\! \left(\sum\limits_{i=1}^{R}\sum\limits_{k=0}^{[\frac{i+1}{2}]}\gamma_{i}2^{i+1-2k}\dbinom{i+1-k}{k}\dfrac{q_{N}^{i+1-2k}}{\left(\sum\limits_{j=1}^{N-1}q_{j}^{2}\right)^{i+2-k}}
\right)\,(M>R).\notag\\
\end{eqnarray}


\sect{Further generalizations}

Summarizing, the main result contained in the previous pages is the fact that the multiparametric ND Hamiltonian 
\be
\mathcal{H}^{(N)}=\dfrac{1}{2}\sum\limits_{i=1}^{N-1}\left(
p_{i}^{2}+\dfrac{\otra_{i}}{q_{i}^{2}}
\right)+\dfrac{1}{2}p_{N}^{2}+\mathcal{V}\left(\sum\limits_{i=1}^{N-1}q_{i}^{2},q_{N}\right)
\label{hgen}
\ee
is completely integrable provided that the function $\mathcal{V}$ is any of the 2D integrable HH potentials. Moreover, the full set of integrals of the motion for (\ref{hgen}) can be explicitly obtained as a consequence of its underlying $sl(2,\mathbb{R}) \oplus h_{3}$ symmetry. 

However, none of the arguments underlying the previous construction prevents us to use it for {\em any} 2D integrable potential of the form $\mathcal{V}\left(x^2,y\right)$ (see the classifications in~\cite{Hindues, 2toN, Hietarinta, Perelomov}). For instance, the 2D Holt potential (see~\cite{2toN, Perelomov})
\be
\mathcal{H}^{(2)}=\dfrac{1}{2}(p_{1}^{2}+p_{2}^{2})+q_{2}^{-\frac{2}{3}}\left(
q_{1}^{2}+\dfrac{9}{2}q_{2}^{2}
\right)
\ee
is also of the form $\mathcal{V}\left(x^2,y\right)$ and its (quartic) integral of the motion is given by
\begin{eqnarray}
\mathcal{I}&=& p_{1}^{4}+2 p_{1}^{2}p_{2}^{2}+24 q_{2}^{\frac{1}{3}}p_{2}q_{1}p_{1}+4q_{2}^{-\frac{2}{3}}q_{1}^{2}p_{1}^{2}+72 q_{2}^{\frac{2}{3}}q_{1}^{2}.
\end{eqnarray}
Therefore, this system admits a $sl(2,\mathbb{R}) \oplus h_{3}$ formulation, that leads to the Hamiltonian
\be
\mathcal{H}=\dfrac{1}{2}(J_{+}+A_{+}^{2})+A_{-}^{-\frac{2}{3}}\left(
J_{-}+\dfrac{9}{2}A_{-}^{2}
\right)
\label{hoh}
\ee
and to the abstract integral
\begin{eqnarray}
\mathcal{I}&=& J_{+}^{2}+2J_{+}A_{-}^{2}+24 A_{-}^{\frac{1}{3}}A_{+}J_{3}+4A_{-}^{-\frac{2}{3}}J_{+}J_{-}+72 A_{-}^{\frac{2}{3}}J_{-}.
\label{hoi}
\end{eqnarray}
As a consequence, by substituting (\ref{sympbi})--(\ref{sh6bi}) in (\ref{hoh}) and (\ref{hoi}) we get an ND integrable generalization of the Holt potential, for which the remaining integrals are again given by the `universal' set (\ref{sl2casbi}). As a side remark, in case that the additional integral $\cal I$ would be unknown, note that this algebraic approach allows the efficient use of symbolic computation packages in order to look for such a constant of the motion, since the ND Hamiltonian can always be written as an object defined on the set of six generators that generate the Poisson algebra $sl(2, \mathbb{R})\oplus h_3$. 

Besides its immediate applicability to many other Hamiltonians, the Poisson symmetry framework here presented opens the path to several further developments on HH systems based on such algebraic perspective, that we plan to work out in the near future. Firstly, the quantization of all the integrable HH systems here presented can be faced by considering the Lie algebra $sl(2,\mathbb{R}) \oplus h_{3}$ as the corresponding underlying symmetry (regarding the many facets of  the quantization problem for the original HH system (\ref{HHaut}) see~\cite{Gutzwiller}). Therefore, the explicit solution of these quantum HH systems should be amenable by making use of the representation theory of this algebra and by diagonalizing the Hamiltonian $\hat{\mathcal{H}}$ in a common eigenbasis with respect to the (suitable defined) quantum `abstract' invariant $\hat{\mathcal{I}}$.

Secondly, the type of Poisson approach here used should be the suitable one in order to define appropriate (integrable) analogues of the HH systems on {\em curved spaces}, that should lead to the flat HH Hamiltonians when the curvature parameter is turned off. 
In this respect, we have recently presented in~\cite{Annals} a constructive $sl(2,\mathbb{R})$-Poisson algebra approach to many integrable systems defined on ND conformally flat spaces. 
In particular, the problem of getting a consistent definition of integrable HH systems on the spherical and hyperbolic spaces should be guided by the preservation of some underlying symmetries that should be related to the flat $sl(2,\mathbb{R}) \oplus h_{3}$ ones, and constitutes a challenging problem. 

Also, it is well known that one of the main features of the Hamiltonian systems endowed with a given Poisson coalgebra symmetry is the fact that integrable {\em deformations} of them can be systematically constructed by introducing a suitable $q$-deformation of the underlying coalgebra (see~\cite{BR,  CRMAngel, RutwigProc}). In our case, one could assume the same `abstract' form of the HH Hamiltonians here considered but -for instance- with the functions $(J_{+},J_{-},J_{3})$ being now generators of a given $q$-deformation of $sl(2,\mathbb{R})$. In that case, the quantum group theory tells us how to deform the full set of integrals (\ref{sl2casbi}), and the only remaining task would be to find the appropriate deformation of the integral ${\mathcal{I}}$. Previous experience on the subject shows that integrable systems on curved spaces can arise in this way, and in such cases the deformation parameter is directly related with the curvature of the underlying space~\cite{PLB}.

Finally, we recall that (1+1) Minkowskian HH systems with kinetic energy of the type $(p_1^2-p_2^2)$ have been considered in the literature in connection with geodesics in gravitational waves~\cite{geodesics}. Indeed, the integrability properties of this class of HH systems could be analysed in terms of the $sl(2,\mathbb{R}) \oplus h_{3}$ symmetry, thus allowing their generalization to an (N+1) Minkowskian space by following the approach here presented. Another interesting problem in a General Relativity context seems to be the study of the algebraic integrability properties of HH systems perturbed by a Kepler potential~\cite{keplerp}, a term that in our language would be written as a perturbation of the type $1/\sqrt{J_- + A_-^2}$.


\section*{Acknowledgements}


This work was partially supported by the Spanish `Ministerio de Ciencia e Innovaci\'on' under grants MTM2007-67389 (with EU-FEDER support) and by Junta de Castilla y
Le\'on  (project GR224).




\begin{thebibliography}{99}

\bibitem{HH} M. H\'enon, C. Heiles, Astron.  J.
\textbf{69}, 73 (1964).

\bibitem{Tabor} M. Tabor, {\it Chaos and integrability in nonlinear dynamics}, (Wiley: new York) (1989).

\bibitem{Gutzwiller} M.C. Gutzwiller,  {\it Chaos in Classical and Quantum
Mechanics}, Interdisciplinary Applied Mathematics, Vol.
1 (Springer: New York) (1990).

\bibitem{BoPu} D. Boccaletti,  G. Pucacco, \textit{Theory of Orbits}, (Berlin: Springer) 
(2004).


\bibitem{BSV} T. Bountis, H. Segur, F. Vivaldi, Phys. Rev. A
\textbf{25},  1257 (1982).

\bibitem{CTW} Y.F. Chang, M. Tabor, J. Weiss,
J. Math. Phys. \textbf{23}, 531 (1982).

\bibitem{GDP} B. Grammaticos,  B. Dorizzi, R. Padjen, Phys. Lett. A \textbf{89},  111 (1982).

\bibitem{HietarintaRapid} J. Hietarinta,  Phys. Rev. A
{\bf 28}, 3670 (1983).

\bibitem{Fordy83} A.P. Fordy,
Phys. Lett. A \textbf{97}, 21 (1983).

\bibitem{Wojc} S. Wojciechowski,
Phys. Lett. A \textbf{100}, 277 (1984).

\bibitem{SL} R. Sahadevan, M. Lakshmanan,  J. Phys. A: Math. Gen. {\bf 19}, L949 (1986).

\bibitem{FordyHH} A.P. Fordy,
Physica D \textbf{52}, 204 (1990).

\bibitem{Sarlet} W. Sarlet,  J. Phys. A: Math. Gen. {\bf 24}, 5245 (1991).

\bibitem{RGG} V. Ravoson, L. Gavrilov, R. Gabor, J. Math. Phys.
\textbf{34}, 2385 (1993).

\bibitem{Hindues} M. Lakshmanan, R. Sahadevan,  Phys. Rep. {\bf 224}, 1 \& 2 (1993).

\bibitem{Pickering} R. Conte, A.P. Fordy, A. Pickering,
Physica D \textbf{69}, 33 (1993).

\bibitem{salerno} M. Salerno, V.Z. Enol'skii, D.V. Leykin, Phys. Rev.
E \textbf{49}, 5897 (1994).

\bibitem{VMC} C. Verhoeven, M. Musette, R. Conte, J. Math. Phys. \textbf{43}, 1906 (2002).

\bibitem{KGM} N. A. Kostov, V.S. Gerdjikov, V. Mioc,  J. Math. Phys. \textbf{51}, 022702 (2010)

\bibitem{2toN} B. Grammaticos, B. Dorizzi, A. Ramani,  J. Hietarinta,
Phys. Lett. A \textbf{109}, 81 (1985).

\bibitem{RauchHH} M. Antonowicz, S. Rauch-Wojciechowski,
Phys. Lett. A \textbf{163}, 167 (1992).

\bibitem{HoneIP}
A.N.W. Hone, V. Novikov, C. Verhoeven, Inv. Probl. {\bf 22}, 2001 
(2006).

\bibitem{HonePLA}
A.N.W. Hone, V. Novikov, C. Verhoeven, Phys. Lett. A {\bf 372}, 1440  (2008).

\bibitem{GrammaticosExtender} B. Grammaticos, A. Ramani,
Phys. Lett. A \textbf{139}, 299 (1989).

\bibitem{BR} A. Ballesteros, O. Ragnisco, J. Phys. A: Math. Gen. {\bf 31}, 3791 (1998).

 \bibitem{CRMAngel} A. Ballesteros, F.J. Herranz,  F. Musso,  O. Ragnisco.
 {\em Superintegrability in Classical and Quantum
Systems} {CRM Proceedings and Lecture Notes} {\bf 37} ed P Tempesta {\em
et al}
(Providence, RI: AMS), 1 (2004).

\bibitem{alfonso} A. Ballesteros, A. Blasco,
J. Phys. A: Math. Theor. \textbf{41}, 304028
(2008).

\bibitem{RutwigProc}
A. Ballesteros, A. Blasco, F.J. Herranz,  F. Musso,  O. Ragnisco, J. Phys: Conf. Ser. \textbf{175}, 012004
(2009).

\bibitem{tesis}
A. Blasco, \textit{Integrability of non-linear Hamiltonian systems with $N$ degrees of freedom}, PhD. Thesis, Burgos University (2009)

\bibitem{Annals} A. Ballesteros, A. Enciso, F.J. Herranz, O. Ragnisco,
Ann. Phys. \textbf{324}, 1219
(2009).

\bibitem{Conte} R. Conte, M. Musette, C. Verhoeven, Theor. Math. Phys., \textbf{144}, 888 (2005).

\bibitem{GDRII} B. Grammaticos, B. Dorizzi, A. Ramani, J. Math. Phys. \textbf{25}, 12 (1984).

\bibitem{Hietarinta} J. Hietarinta, Phys. Rep. \textbf{147}, 87 (1987).

\bibitem{Perelomov} A.M. Perelomov. \textit{Integrable systems of classical mechanics and Lie algebras}, (Basel: Birh$\ddot{\text{a}}$userVerlag) (1990).

\bibitem{Tondo} G. Tondo,
 J. Phys. A: Math. Gen. \textbf{28}, 5097 (1995).

\bibitem{RDGprl}
A. Ramani, B. Dorizzi, B. Grammaticos, Phys. Rev. Lett. {\bf 49}, 1539  (1982).

\bibitem{KCJPA} A. Ballesteros, F.J. Herranz,
J. Phys. A: Math. Theor. \textbf{42}, 245203 (2009).

\bibitem{PLB}
A.   Ballesteros, A. Enciso, F.J. Herranz, O. Ragnisco, {Phys. Lett. B}  {652}, 376 (2007).

\bibitem{geodesics} K. Vesely, J. Podolsky, {Phys. Lett. A}  {271}, 368 (2000).

\bibitem{keplerp} F. Kokubun, {Phys. Lett. A}  {245},  358 (1998).





\end{thebibliography}
\end{document}